\documentclass{PoS}

\title{New results in exclusive hard reactions}

\ShortTitle{New results in exclusive hard reactions}

\author{\speaker{B. Pire}\\
        CPhT, \'Ecole Polytechnique, CNRS, 91128 Palaiseau, France\\
      E-mail: \email{pire@cpht.polytechnique.fr}
}
\author{K.~Semenov-Tian-Shansky\\
 CPhT, \'Ecole Polytechnique, CNRS, 91128 Palaiseau, France \\ LPT, Universit\'e Paris-Sud, CNRS, 91405 Orsay, France}
\author{L. Szymanowski and J. Wagner
\\NCNR, Warsaw, Poland\\}

\abstract{Generalized Parton Distributions offer a new way to access the quark and gluon nucleon structure. We review recent progress in this domain, emphasizing the need to supplement the experimental study of DVCS by its crossed version, timelike Compton scattering (TCS), where data at high energy should appear thanks to the study of ultraperipheral collisions at the LHC. This will open the access to very low skewness quark and gluon GPDs. Our leading  order estimates show that the factorization scale dependence of the amplitudes is quite high. This fact demands the understanding of higher order contributions with the hope that they will stabilize this scale dependence. The magnitudes of the NLO coefficient functions are not small and neither is the  difference of the coefficient functions  appearing respectively in the DVCS and TCS amplitudes. The conclusion is that extracting the universal GPDs from both TCS and DVCS reactions requires much care. We also describe the extension of the GPD concept to three quark operators and the relevance of their nucleon to meson matrix elements, namely the transition distribution amplitudes (TDAs) which factorize in hard exclusive pion electroproduction off a nucleon in the backward region and baryon-antibaryon annihilation into a pion and a lepton pair. We discuss the main properties of the TDAs. }

\FullConference{The 2011 Europhysics Conference on High Energy Physics-HEP 2011,\\
		July 21-27, 2011\\
		Grenoble, Rh\^one-Alpes, France}

\begin{document}

The study of the deep structure of the nucleon has been the subject of many developments in the past decades and the concept of generalized parton distributions has allowed a breakthrough in the 3 dimensional description  of the quark and gluon content of hadrons. Hard exclusive reactions have been demonstrated to allow to probe the 
quark and gluon content of protons and heavier nuclei.
A considerable amount of theoretical and experimental work has 
been devoted to the study of deeply virtual Compton scattering (DVCS),
 i.e., $\gamma^* p \to \gamma p$, 
an exclusive reaction where generalized parton
distributions (GPDs) and perturbatively calculable coefficient functions factorize in the scattering amplitude, when
the virtuality of the incoming photon is high enough.
It is now recognized that the measurement of GPDs should contribute in a decisive way to
our understanding of how quarks and gluons assemble into
hadrons~\cite{gpdrev}. In particular the transverse
location of quarks and gluons become experimentally measurable via the transverse momentum dependence of the GPDs \cite{Burk}. Results on DVCS obtained at HERA and JLab already allow to get a rough 
idea of some of the GPDs (more precisely on the Compton form factors) in a restricted  kinematical domain \cite{fitting}. An extended research program at JLab@12 GeV and 
Compass is now proposed to go beyond this first set of analysis. 
This will involve taking into account  next to leading order in $\alpha_s$ and higher twist contributions \cite{APT}.

\section{On Timelike Compton scattering}
The physical way to observe the inverse reaction,  timelike Compton scattering (TCS) \cite{BDP},
 \begin{eqnarray}
 \gamma(q) N(p) \to \gamma^*(q') N(p') \nonumber
 \end{eqnarray}
 is  the exclusive photoproduction of a
heavy lepton pair, $\gamma N \to \mu^+\!\mu^-\, N$ or $\gamma N \to
e^+\!e^-\, N$, at small $t = (p'-p)^2$ and large \emph{timelike} invariant squared mass of the final state lepton pair  $q'^2 = Q'^2$. TCS 
shares many features with DVCS. The generalized Bjorken variable is $\tau = Q'^2/s $
 with $s=(p+q)^2$. One also defines $\Delta = p' -p$ , $t= \Delta^2$ and the skewness variable 
$\eta = - \frac{(q-q')\cdot (q+q')}{(p+p')\cdot (q+q')} \,\approx\,
           \frac{ Q'^2}{2s  - Q'^2} = \frac{ \tau}{ 2-\tau}.$
 At the Born order, the TCS amplitude is described by the usual handbag diagrams.
As in the case of DVCS,  a purely electromagnetic
mechanism where the lepton pair is produced through the Bethe-Heitler (BH) subprocess 
$\gamma (q)\gamma^* (\Delta) \to\ell^+\ell^-\;,$  contributes at the amplitude level. 
Since the amplitudes for the Compton and Bethe-Heitler
processes transform with opposite signs under reversal of the lepton
charge,  the interference term between TCS and BH is
odd under exchange of the $\ell^+$ and $\ell^-$ momenta.
It is thus possible to
 project out
the interference term through a clever use of
 the angular distribution of the lepton pair. 
The interference part of the unpolarized cross-section for $\gamma p\to \ell^+\ell^-\, p$  has a characteristic ($\theta, \varphi$) dependence given  by (see details in \cite{PSW})
\begin{eqnarray}
   \label{intres}
\frac{d \sigma_{INT}}{dQ'^2\, dt\, d\cos\theta\, d\varphi}
= {}-
\frac{\alpha^3_{em}}{4\pi s^2}\, \frac{1}{-t}\, \frac{M}{Q'}\,
\frac{1}{\tau \sqrt{1-\tau}}\,
  \cos\varphi \frac{1+\cos^2\theta}{\sin\theta}
    {\cal R}e{\cal M} \; ,\nonumber
\end{eqnarray}
with 
\begin{eqnarray}
\label{mmimi}
{\cal M} = \frac{2\sqrt{t_{min}-t}}{M}\, \frac{1-\eta}{1+\eta}\,
\left[ F_1 {\cal H} - \eta (F_1+F_2)\, \tilde{\cal H} -
\frac{t}{4M^2} \, F_2\, {\cal E} \,\right], \nonumber
\end{eqnarray}
where  ${\cal H},  \tilde{\cal H}, {\cal E}$ are the Compton form factors, i.e. the convolutions of GPDs and coefficient functions, and $F_1, F_2$ are the nucleon Dirac and Pauli form factors.
The interference
term changes sign under $\varphi\to \pi+\varphi$ due to charge conjugation,
whereas the TCS and BH cross sections do not. One may thus extract the 
Compton amplitude through a study of
$\int\limits_0^{2\pi}d\phi\,\cos \phi \frac{d\sigma}{d\phi}$  provided  the  $\theta$ integration is performed within limits symmetric about $\theta=\pi/2$ .

 This program has not yet been experimentally successful due to the existing limited quasi real photon flux in the right kinematical domain both at JLab and HERA. This will be  much improved with the JLab@12 GeV program. Experiments at higher energies, e.g. in ultraperipheral collisions at RHIC and LHC \cite{PSW}, may even become sensitive to gluon GPDs which enter the amplitude only at NLO level.
TCS and DVCS amplitudes are identical (up to a complex conjugation) at lowest order in $\alpha_S$ but differ at next to leading order.
Indeed the production of a timelike photon enables the production of intermediate states in some channels which were kinematically forbidden in the DVCS case. This opens the way to new absorptive parts of the amplitude.  We have demonstrated \cite{PSW2} that these NLO differences are quite important and cannot be neglected. This shows that experiments  will enable to test the universality of GPDs extracted from DVCS and from TCS, only if  NLO corrections are properly taken into account.

\section{On Transition Distribution Amplitudes}
 
The factorization theorem for backward DVCS \cite{PS} and for hard exclusive backward meson electroproduction argued in
\cite{Frankfurt:1999fp}
lead to the introduction of baryon to meson transition distribution amplitudes (TDAs),
non diagonal matrix elements
of light-cone three quark operators
$
\widehat{O}^{\alpha \beta \gamma}_{\rho \tau \chi}(z_1,\,z_2,z_3)=
\varepsilon_{c_1 c_2 c_3}
\Psi^{c_1 \, \alpha}_\rho(z_1) \Psi^{c_2 \beta}_\tau (z_2) \Psi^{c_3 \, \gamma}_\chi (z_3)\nonumber
$
between baryon and meson states. Here, $\alpha$, $\beta$, $\gamma$
stand for quark flavor indices;
$\rho$, $\tau$
and
$\chi$
denote the Dirac indices and
$c_{1,2,3}$
are indices of the color group.
If one adopts the light-cone
gauge
$A^+=0$,
 the gauge link is equal to unity and may be omitted
in the definition of the operator.
Baryon to meson Transition Distribution Amplitudes  extend
the concept of generalized parton distributions. They appear
as a building block in the colinear factorized description of amplitudes for a class of
hard exclusive reactions prominent examples being hard exclusive pion electroproduction off a nucleon
in the backward region and baryon-antibaryon annihilation into a pion and a lepton pair \cite{Pire:2005ax,LPS} which may be studied with PANDA at GSI-FAIR~\cite{Lutz:2009ff}.
The definition of leading twist-$3$ $\pi N$ TDAs can be symbolically written as (see details in  \cite{Pire:2011xv}):
\begin{eqnarray}
&&
4(P \cdot n)^3 \int   \left[ \prod_{j=1}^3 \frac{d \lambda_j}{2 \pi}   \right]
e^{i \sum_{k=1}^3 x_k \lambda_k (P \cdot n)}
\langle \pi_a(p_\pi)| \widehat{O}_{\rho \, \tau \, \chi}^{\alpha \beta \gamma}( \lambda_1 n, \,\lambda_2 n, \, \lambda_3 n )| N_\iota(p_1)  \rangle
\nonumber \\ &&
=\delta(x_1+x_2+x_3-2 \xi) \; \sum_{s.f.} (t_a)^{\alpha \beta \gamma}_\iota s_{\rho \, \tau \, \chi} H^{(\pi N)}_{s.f.}(x_1,x_2,x_3,\xi, \Delta^2)\,,\nonumber
\label{Formal_definition_TDA}
\end{eqnarray}
where the spin-flavor ($s.f.$) sum  stands over all relevant independent
flavor structures $(t_a)^{\alpha \beta \gamma}_\iota$
and the Dirac structures
$s_{\rho \, \tau \, \chi}$.  The skewness parameter $\xi$ is defined similarly as in the
GPD case by $\xi=- \frac{\Delta \cdot n}{2 P \cdot n}$, $\Delta$ referring to the $u-$channel transfer..

$\pi N$ TDAs are non-perturbative objects governed by long distance dynamics.  
In accordance with the usual logic of the collinear factorization approach, they satisfy 
 well established renormalization group equations.
The evolution properties of the three quark non-local operator on the light-cone
were extensively studied in the literature
for the case of matrix elements
between a baryon and the vacuum known as baryon distribution amplitudes (DAs). Since TDAs involve the same operator, their  evolution is also determined. 
The Lorentz invariance results in
a polynomiality property of the Mellin moments of TDAs in the longitudinal
momentum fractions. We derived \cite{Pire:2010if} a spectral representation
for the $\pi N$ TDAs, and introduced the notion of quadruple
distributions. This enables us to generalize
Radyushkin's factorized Ansatz from the GPD case to the TDA case.

The detailed account of isospin and permutation symmetries \cite{Pire:2011xv}
provides a unified description of all isotopic channels in terms of eight 
independent $\pi N$ TDAs. 
These general constraints derived  should be satisfied by 
any realistic model  of TDAs. 
The crossing relation between $\pi N$ TDAs and GDAs leads to 
a soft pion theorem for isospin-$\frac{1}{2}$ and isospin-$\frac{3}{2}$ $\pi N$ TDAs, which helps to derive normalization conditions for $\pi N$ TDAs. 
A simple resonance exchange model considering nucleon
and $\Delta(1232)$ exchanges in 
isospin-$\frac{1}{2}$ and isospin-$\frac{3}{2}$ channels respectively allows  to approximate $\pi N$ TDAs in the ERBL region \cite{Pire:2011xv}. Nucleon exchange 
may be considered as a pure $D$-term contribution generating 
the highest power monomials of $\xi$ of the Mellin moments
in the longitudinal momentum fractions,  complementary to the spectral representation
for TDAs in terms of quadruple distributions. 
We can now write down consistent models of TDAs to be confronted with experimental data, which should become available in the coming years, both from JLab at 12 GeV and from PANDA at FAIR.

\vskip.1in
\noindent
 {\bf Acknowledgments}

\noindent 
This work is partly supported by the Consortium Physique des Deux Infinis (P2I).

\end{document}